\documentclass[10pt, conference]{IEEEtran}
\IEEEoverridecommandlockouts
\usepackage{cite}
\usepackage{amsmath,amssymb,amsfonts}
\usepackage{bbm}
\usepackage{algorithm,algorithmic}
\usepackage{graphicx}
\usepackage{textcomp}
\usepackage{xcolor}
\usepackage{soul}
\usepackage{siunitx}
\usepackage[printonlyused]{acronym}
\usepackage{orcidlink}
\usepackage{gensymb}
\usepackage{makecell}
\usepackage{tikz}
\setcellgapes{1pt}

\acrodef{mmW}{millimeter-wave}
\acrodef{sub-THz}{sub-terahertz}
\acrodef{BS}{base station}
\acrodef{UE}{user equipment}
\acrodefplural{UE}{user equipment}
\acrodef{Tx}{transmitter}
\acrodef{Rx}{receiver}
\acrodef{SNR}{signal-to-noise ratio}
\acrodef{AWV}{antenna weight vector}
\acrodef{LOS}{line-of-sight}
\acrodef{NLOS}{non-line-of-sight}
\acrodef{AWGN}{additive Gaussian white noise}

\acrodef{CS}{compressive sensing}
\acrodef{CPR}{compressive phase retrieval}
\acrodef{BA}{beam alignment}
\acrodef{BT}{beam tracking}
\acrodef{BP}{beam prediction}
\acrodef{IA}{initial access}
\acrodef{AoA}{angles-of-arrival}
\acrodef{AoD}{angle-of-departure}
\acrodef{RSS}{received signal strength}
\acrodef{MP}{matching pursuit}
\acrodef{ML}{machine learning}
\acrodef{PN}{pseudorandom noise}
\acrodef{MC}{Monte Carlo}

\acrodef{SA}{sub-array}
\acrodef{PA}{phased array}
\acrodefplural{PA}[PAs]{phased arrays}
\acrodef{TTD}{true-time delay}
\acrodef{PS}{phase shifter}
\acrodef{JPTA}{joint phase-time array}
\acrodef{OFDMA}{orthogonal frequency-division multiple access}
\acrodef{QPD}{quadratic phase distribution}

\acrodef{DL}{downlink}
\acrodef{ICI}{intercarrier interference}
\acrodef{CFO}{carrier frequency offset}
\acrodef{IUI}{inter-user interference}
\acrodef{SINR}{signal-to-interference-and-noise ratio}
\acrodef{CDF}{cumulative density function}

\acrodef{URLLC}{ultra-reliable and low-latency communication}
\acrodef{V2X}{vehicle-to-everything}
\acrodef{IoT}{Internet-of-Things}
\acrodef{AR}{augmented reality}
\acrodef{VR}{virtual reality}
\acrodef{XR}{extended reality}

\newcommand\copyrighttext{%
  \footnotesize \textcopyright 2025 IEEE. Personal use of this material is permitted.
  Permission from IEEE must be obtained for all other uses, in any current or future
  media, including reprinting/republishing this material for advertising or promotional
  purposes, creating new collective works, for resale or redistribution to servers or
  lists, or reuse of any copyrighted component of this work in other works. 
}
\newcommand\copyrightnotice{%
\begin{tikzpicture}[remember picture,overlay]
\node[anchor=south,yshift=10pt] at (current page.south) {\fbox{\parbox{\dimexpr\textwidth-\fboxsep-\fboxrule\relax}{\copyrighttext}}};
\end{tikzpicture}%
}

\def\BibTeX{{\rm B\kern-.05em{\sc i\kern-.025em b}\kern-.08em
    T\kern-.1667em\lower.7ex\hbox{E}\kern-.125emX}}

\begin{document}
\bstctlcite{IEEEexample:BSTcontrol}

%%%%%%%%%%%%%%%%%%%%%%%%%%%%%%%%%%%%%%%%%%%%%%%%%%%%%%%
%%%%%%%%% Actual Paper Content
%%%%%%%%%%%%%%%%%%%%%%%%%%%%%%%%%%%%%%%%%%%%%%%%%%%%%%%

\title{Millimeter-Wave True-Time Delay Array Beamforming with Robustness to Mobility \\
\thanks{This material is based upon work supported by the National Science Foundation under Grant No. 2224322.}
}

\author{\IEEEauthorblockN{Benjamin W. Domae, Ibrahim Pehlivan, and Danijela Cabric	}
		
\IEEEauthorblockA{\textit{Electrical and Computer Engineering Department,} \\
\textit{University of California, Los Angeles}\\
%City, Country \\
Emails: bdomae@ucla.edu, ipehlivan@ucla.edu, danijela@ee.ucla.edu }
}

\maketitle
\copyrightnotice

\begin{abstract}

Ultra-reliable and low-latency connectivity is required for real-time and latency-sensitive applications, like wireless augmented and virtual reality streaming. 
Millimeter-wave (mmW) networks have enabled extremely high data rates through large available bandwidths but struggle to maintain continuous connectivity with mobile users. Achieving the required beamforming gain from large antenna arrays with minimal disruption is particularly challenging with fast-moving users and practical analog mmW array architectures. In this work, we propose frequency-dependent slanted beams from true-time delay (TTD) analog arrays to achieve robust beamforming in wideband, multi-user downlink scenarios. Novel beams with linear angle-frequency relationships for different users and sub-bands provide a trade-off between instantaneous capacity and angular coverage. Compared to alternative analog array beamforming designs, slanted beams provide higher reliability to angle offsets and greater adaptability to varied user movement statistics.

\end{abstract}

\begin{IEEEkeywords}
URLLC, beamforming, millimeter-wave, true-time delay array, joint phase time array
\end{IEEEkeywords}

\section{Introduction}

Future 6G cellular communications networks are developing \ac{URLLC} capabilities to support applications that require high-reliability. 
\Ac{XR}, in particular, requires high data rates with extremely low latency and high reliability, as high latency bursts and connectivity disruptions can cause discomfort for users \cite{Marinsek2024_urllc_vr}.
\Ac{mmW} and \ac{sub-THz} systems, operating at 30-300 GHz, can meet this demand for high data rates but must properly align narrow antenna array beams to maintain sufficient \ac{SNR} for communications. 
For \ac{XR}, consistent \ac{BA} is especially critical due to rapid user movement and rotation. 
Current \ac{mmW} devices complete this \ac{BA} over time using a periodic exhaustive search of all potential beams. However, these searches are undesirable for \ac{URLLC}, as their latency scales linearly with array size due to narrower beams.

One way to reduce \ac{BA} delay is to conduct \ac{BA} less frequently. \cite{Jayaprakasam2017_tracking_kf} and \cite{Domae2022_track_lstm} increase the \ac{BA} repetition period by predicting beams with Kalman filtering or machine learning respectively. This strategy reduces average \ac{BA} delays over time but can leave gaps in coverage due to tracking inaccuracy or beam switching delays from array hardware. 
Wider beams allow for less frequent beam updates by trading some beamforming gain for more angular coverage, enabling the same beam to be used for more time where a user may be moving. Increased angular coverage increases the beam's \textit{robustness}, or resiliency to unknown angle offsets from movement or errors. Prior algorithms have utilized wider beams from smaller arrays \cite{Va2016_wide_track}, arrays with deactivated elements \cite{Chung2021_wide_adaptive}, or simplified models \cite{Zhou2022_wide_cb} for more efficient tracking. \cite{Chung2021_wide_adaptive} and \cite{Zhou2022_wide_cb} each additionally optimized beamwidths for robustness to angle offsets.

\begin{figure*}[t]
    \vspace{0mm}
    \centering
    % left bottom right top
    \includegraphics[width = 0.94\linewidth,trim=15 20 15 15,clip]{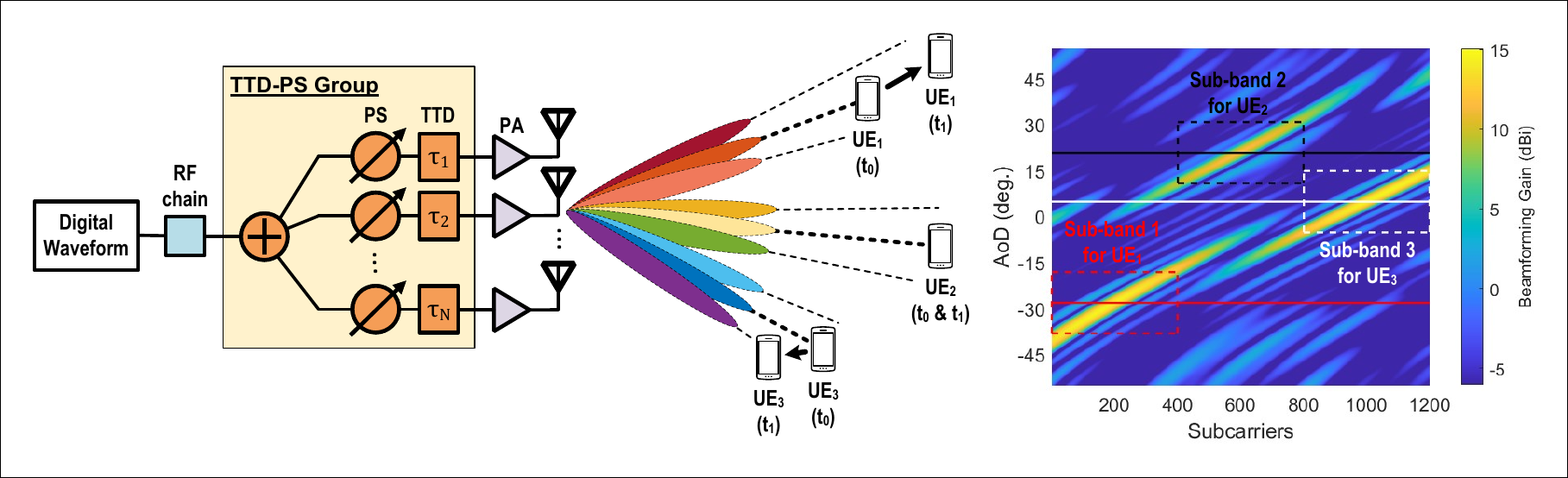}
    % \vspace{-6mm}
    \vspace{-3mm}
    \caption{System architecture with three users (and three sub-bands) with a corresponding example slanted beams pattern.}
    \label{fig:system}
    \vspace{-5mm}
\end{figure*}

\Ac{TTD} arrays, also known as \acp{JPTA}, are simple analog \acp{PA} arrays with added \ac{TTD} elements that enable more flexible beamforming and faster \ac{BA}. With a single RF chain, \ac{TTD} arrays can control the frequency-dependence of their beam patterns \cite{Boljanovic2021_ttd}. In wideband scenarios, \ac{PA} beams are also frequency-dependent from beam squint, but this angle-frequency relationship is not controllable. While fully-digital and hybrid arrays may also have frequency-dependent beamforming capability, their multiple RF chains can be expensive and power hungry. 
Several recent works have shown \ac{TTD} arrays' promise for flexible wideband beamforming for \ac{OFDMA} and fast \ac{BA} \cite{Yan2019_ttd_rainbow, Ratnam2022_ttd_jpta, Zhao2024_ttd_step}. \ac{TTD} arrays have also been studied for robustness to mobility with frequency-stepped \cite{Mo2023_ttd_mobile} and sector-dispersive \cite{Zhou2024_ttd_track} beams, but these methods are most efficient for single user cases. 

In this work, we propose a robust downlink beamforming design using functionally-wider beams from analog-TTD arrays. Our \textit{slanted beams} solution creates beams with approximately linear \ac{AoD}-to frequency relationships, like \cite{Yan2019_ttd_rainbow} and \cite{Zhou2024_ttd_track}, but over specified angle sectors associated with contiguous sub-bands for multi-user \ac{OFDMA}. Slanted beams provide a way to trade-off capacity per user for mobility robustness; steeper slopes in the \ac{AoD}-frequency domain cover more \acp{AoD} but have fewer strong subcarriers in each sub-band. We design the angular range to meet specified robustness requirements, providing a reliable but heavily frequency-faded channel over each user's sub-band.

The remainder of this work is organized as follows: Section \ref{sec:system} discusses our system model, problem statement, and metrics. Section \ref{sec:beams} details the selected baseline beamforming and our proposed slanted beams design, while Section \ref{sec:results} presents our simulation results comparing slanted beams to the baseline beams. Finally, we conclude in Section \ref{sec:conclusion}.

\textit{Notation}: Scalars, vectors, and matrices are represented by non-bold, bold lowercase, and bold uppercase letters respectively. The $i$th entry of $\mathbf{a}$ is denoted as $\mathbf{a}_i$. 
For $\mathbf{A}$, the $i$th row is $[A]_{i,:}$ and the $j$th element of the $i$th row is $[\mathbf{A}]_{i,\text{j}}$. 
The transpose and Hermitian transpose of $\mathbf{A}$ are $\mathbf{A}^T$ and $\mathbf{A}^H$ respectively.

\section{System Model and Problem Statement}
\label{sec:system}

We consider a \ac{URLLC} \ac{DL} \ac{mmW} \ac{OFDMA} system, where a \ac{BS} beamforms towards each potentially-moving \ac{UE} and sends each \ac{UE}'s continuous data stream on separate sub-bands. While higher data rates are better, our primary objective is to maintain a highly-reliable minimum data rate at all times for all \acp{UE}. An overview of our system is shown in Fig. \ref{fig:system} with three users.

\subsection{True-Time Delay Array Beamforming and SINR}
\label{subsec:system_ttd}

The analog \ac{TTD} array architecture is key to our unique slanted beams. Fig. \ref{fig:system} shows the \ac{BS} \ac{Tx} \ac{TTD} array architecture. Each antenna element has one \ac{TTD} and one \ac{PS}, enabling controlled beam squint through the combination of frequency-independent time delays and frequency-dependent phase shifts. 
The \ac{TTD} beamforming \ac{AWV} for a vector of delays $\mathbf{\tau} \in \mathbb{R}^{N_t}$ and vector of phases $\mathbf{\phi} \in \mathbb{R}^{N_t}$ can expressed for antenna element $n$ and frequency $f$ as \eqref{eq:ttd_bf}, where $N_t$ is the number of \ac{BS} antennas. The gain pattern for this \ac{AWV} at \ac{AoD} $\theta$ and frequency $f$ is then shown in \eqref{eq:ttd_pattern} with $d$-wavelength antenna spacing, center frequency $f_c$, speed of light $c$, and array response $\mathbf{a} \in \mathbb{C}^{N_t}, \mathbf{a}_n( \theta, f ) = \exp{\left( j2\pi(n-1)\sin(\theta) f_c d / c \right)}$. 
\begin{align}
    \label{eq:ttd_bf}
    \mathbf{v}_n\left(f\right) &= \exp{\left(j\left(\mathbf{\phi}_n - 2\pi\mathbf{\tau}_n f\right)\right)} \\
    \label{eq:ttd_pattern}
    a\left(\theta, f\right) &= \left| \mathbf{a}^H\left(\theta, f\right) \mathbf{v}\left(f\right) \right|^2 \\
    \label{eq:ttd_system}
    \mathbf{y}\left(u, t_i\right) &= [\mathbf{H}_u \widetilde{\mathbf{P}}_i \mathbf{x}]_{\kappa_u} + n_u\left[i\right] \\
    \label{eq:ofdma_channel}
    \left[\mathbf{H}_u\left(t_i\right)\right]_{k,k} &= h_u \mathbf{a}^H\left(\theta_u\left(t_i\right), f_k\right) \mathbf{v}\left(f_k\right)     
\end{align}

For \ac{UE} $u$ at \ac{AoD} $\theta_u$ and time step $t_i$, the $K/U$ received symbols from the \ac{TTD} \ac{BS} are described in \eqref{eq:ttd_system} without frequency offsets, where $n_u\left[i\right]$ is \ac{AWGN} with variance $\sigma_N^2$. 
We assume a sequential association of users to sub-bands, so $\kappa_u = \left\{\textstyle\frac{(u-1)K}{U}+1, ..., \textstyle\frac{uK}{U}\right\}$ are the subcarrier indices for user $u$.
\ac{UE} $u$'s channel is assumed to be entirely \ac{LOS}, so $\mathbf{h}_u\left(t_i, f\right) = h_u \mathbf{a}\left(\theta_u\left(t_i\right), f\right)$, where $h_u$ is the channel propagation loss. 
The raw symbols $x \in \mathbb{C}^K$ are normalized to have unit average power, with the \ac{Tx} power allocation is $\mathbf{P} \in \mathbb{R}^{R \times K}$, where $\widetilde{\mathbf{P}}_i = \text{diag}\left(\left[\mathbf{P}_{i,1},...,\mathbf{P}_{i,K}\right]\right)$.

The beamforming frequency fading is given in the $K\times K$ diagonal matrix $\mathbf{H}_u\left(t_i, f\right)$ \eqref{eq:ofdma_channel}.
\ac{OFDMA} subcarrier orthogonality, spatial and frequency filtering from the beamforming, and our \ac{LOS} channel assumption effectively eliminate intercarrier and inter-user interference without frequency offsets. 
The nominal \ac{SNR}, without \ac{Tx} array gain, is $\psi = P_{\text{tot}} / \sigma_N^2$, where $P_{\text{tot}}$ is the total power of the transmitted OFDMA symbol. 
The total \ac{SNR} with \ac{Tx} beamforming is given as $\zeta_{k_u}\left(t_i\right) = |[\mathbf{H}_u \widetilde{\mathbf{P}}_i]_{k_u,k_u}|^2 \textstyle\frac{K}{\sigma_N^2}$, for user $u$'s subcarrier $k_u$.

\subsection{BA Frames and Movement Model}
\label{subsec:system_move}

\begin{figure}[t]
    \vspace{0mm}
    \centering
    % left bottom right top
    \includegraphics[width = 0.95\linewidth,trim=15 22 15 15,clip]{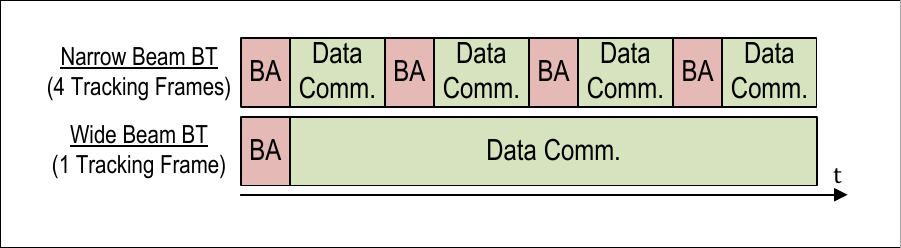}
    \vspace{-3mm}
    \caption{One data communication beam is used per BA frame. With 5G-NR parameters, wide beam tracking frames could be up to 16 times as long as those of narrow beams, requiring fewer BA interruptions.}
    \label{fig:frames}
    \vspace{-5mm}
\end{figure}

In this work, we design a single beamformer to downlink to multiple \ac{UE}s during a single \ac{BA} frame, or a period of $T$ seconds between subsequent \acp{BA}. In 5G New Radio (5G-NR), this is equivalent to designing a single data communications beam for the period $T$ between Synchronization Signal Blocks (SSBs), between 10-160 ms and typically 20 ms \cite{Giordani2019_5g_tracking}. 
As shown in Fig. \ref{fig:frames}, increasing $T$ reduces interruptions to data communications for \ac{BA} overhead. However, higher $T$ could reduce tracking quality, as each \ac{UE} has more time to move outside of the beam.
For our simulations, we break a \ac{BA} frame $T$ into $R$ time steps of length $\Delta t = T / R$.

Our random movement model for user $u$ is shown in \eqref{eq:movement}. Due to the relatively short time scale of each \ac{BA} frame, we simplified each \ac{UE}'s randomized movement over a frame to a one-dimensional angular kinematics model. \ac{UE} $u$'s kinematics, \ac{AoD} $\theta_u(t)$, angular velocity $\omega_u(t)$, and constant angular acceleration $\alpha_u$, are parameterized by random variables for their initial conditions. 
We assume a separate earlier algorithm provides unbiased estimates for $u$'s initial conditions, denoted as $\widehat{\theta}_{u}\left(t_0\right), \widehat{\omega}_{u}\left(t_0\right), \widehat{\alpha}_{u}$. 
We further assume the estimation error is Gaussian-distributed with variances $\sigma_{\theta}$, $\sigma_{\omega}$, and $\sigma_{\alpha}$ respectively. The unknown errors create a Gaussian distribution of $\theta_u\left(t_i\right)$, the possible \ac{AoD}s for each user at each time step, with mean \eqref{eq:move_mean} and variance \eqref{eq:move_var}. Fig. \ref{fig:aod_range} illustrates an example of the density of this distribution.
\begin{align}
    \label{eq:movement}
    \theta_u\left(t_i\right) &= \theta_u\left(t_{0}\right) + \left(i\Delta t\right) \omega_u\left(t_{0}\right) + \left(i\Delta t\right)^2 \textstyle\frac{\alpha_u}{2} \\
    \label{eq:move_mean}
    \widehat{\theta}_{u}\left(t_i\right) &= \widehat{\theta}_u\left(t_{0}\right) + \left(i\Delta t\right) \widehat{\omega}_u\left(t_{0}\right) + \left(i\Delta t\right)^2 \textstyle\frac{\widehat{\alpha}_u}{2} \\
    \label{eq:move_var}
    \sigma_i &= \sigma_{\theta} + \left(i\Delta t\right)^2 \sigma_{\omega} + \left(i\Delta t\right)^4 \textstyle\frac{\sigma_{\alpha}}{4}
\end{align}

\subsection{System Capacity Metric}
\label{subsec:system_metrics}

The primary system metric used in this work is the minimum channel capacity for all users over all time steps. In this work, we approximate a given user's capacity at discrete time steps $t_i$ over a tracking frame using the \ac{Rx} \ac{SNR} at the centers of all subcarriers $f_k$. The user capacity at time step $i$ depends on the \ac{AoD} to the user and the \ac{BS} beamformer $\mathbf{v}$, as described in \eqref{eq:metric_ue_cap}. The subcarrier spacing is $w_{sc} = W/K$, for total system bandwidth $W$ and $K$ subcarriers.
\begin{align}
    \label{eq:metric_ue_cap}
    &C_u\left(t_i, \mathbf{v}\right) = \textstyle\sum_{k=1}^{K} C_{u,k}\left(t_i, \mathbf{v}\right) \\
    &C_{u,k}\left(t_i, \mathbf{v}\right) = w_{sc} \log \left( 1 + \zeta_{k_u} \right) \\
    &=  w_{sc} \log \left( 1 + \mathbf{P}_{i, k} \left|h_u \mathbf{a}^H\left(\theta_u\left(t_i\right), f_k\right) \mathbf{v}\left(f_k\right)\right|^2 \textstyle\frac{1}{\sigma_N^2} \right) \nonumber\\
    &= w_{sc} \log \left( 1 + \mathbf{P}_{i, k} a\left(\theta_u\left(t_i\right), f_k\right) \textstyle\frac{1}{\sigma_N^2} \right)
\end{align}

\subsection{Problem Statement}
\label{subsec:system_problem}

Our objective is to design \ac{BS} \ac{TTD} beams that maximize the minimum capacity given random \ac{UE} movement. This requires the system to maximize the continuously-available capacity for all \ac{UE}s and time steps. This can be expressed as a non-convex optimization problem in \eqref{eq:problem}, with physical constraints on the beamformer $\mathbf{v}$, power allocation $\mathbf{P}$, and maximum \ac{TTD} $\tau_{\text{max}}$.
\begin{align}
    \label{eq:problem}
    \max_{\mathbf{\phi}, \mathbf{\tau}, \mathbf{P}} \min_{u, i} & \hspace{0.5em} C_u\left(t_i, \mathbf{v}\right), \hspace{0.75em} i \in \{1,...,R\}, u \in \{1,...,U\} \\
    \mathrm{s.t.} & \hspace{0.5em} \mathbf{\phi}_n \in [-\pi, \pi], \hspace{0.5em} \mathbf{\tau}_n \in [0, \tau_{\text{max}}] \hspace{0.5em} \forall n \in\{1,...,N_t\} \nonumber \\
    & \hspace{0.5em} \mathbf{P}_{i,k} \geq 0 \hspace{0.5em} \forall i,k \nonumber \\
    & \hspace{0.5em} \textstyle\sum_k^{K} \mathbf{P}_{i,k} = P_{\text{tot}} \hspace{0.5em} \forall i \nonumber
\end{align}

\begin{figure}[t]
    \vspace{0mm}
    \centering
    % left bottom right top
    \includegraphics[width = 0.95\linewidth,trim=110 295 100 295,clip]{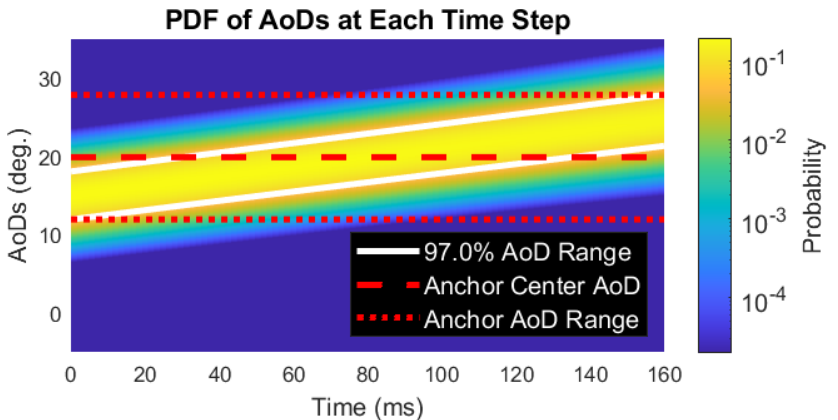}
    \vspace{-3mm}
    % \vspace{-6mm}
    \caption{Example AoD range and anchor selection for one user with $P$=97\%, 
    $\widehat{\theta}_{u}\left(t_{0}\right)$=15\degree, 
    $\widehat{\omega}_{u}\left(t_{0}\right)$=60 deg/s. See Table \ref{table:sims} for the remaining parameters.} 
    % $\widehat{\alpha}$=0 deg\textsuperscript{2}/s\textsuperscript{2},
    % $\sigma_{\theta}$=2 deg\textsuperscript{2}, 
    % $\sigma_{\omega}$=10 deg\textsuperscript{2}/s\textsuperscript{2}, 
    % $\sigma_{\alpha}$=5 deg\textsuperscript{2}/s\textsuperscript{4}.}
    \label{fig:aod_range}
    \vspace{-5mm}
\end{figure}

\section{TTD Slanted Beam Design}
\label{sec:beams}

\subsection{Proposed Slanted Beam Design}
\label{subsec:beams_slanted}

%% General Strategy
The optimization in \eqref{eq:problem} is nontrivial and would require a computationally-complex iterative algorithm to solve directly.
Instead, our proposed slanted beams approximate the solution to \eqref{eq:problem} with a simple heuristic. For large arrays ($N_t \gg 1$), \ac{UE}s at \ac{AoD}s without main-lobe beamforming gain at any subcarriers within their sub-band will have insufficient \ac{SNR}. In other words, to achieve a higher minimum capacity in \eqref{eq:problem}, the designed beamformer must point towards each \ac{UE} for some subcarriers in their respective sub-bands. This simplification of the problem allows us to instead design a heuristic beam that points a fraction of the bandwidth towards every \ac{AoD} from a set of angles that will contain all \ac{UE}s with selected probability $p$. The simplest form of beam pointing directions is a line of directions over the \ac{AoD}-frequency domain of each user's \ac{AoD} range and sub-band. The slope of this line provides a direct trade-off between supported \ac{AoD} range and the number of subcarriers with significant beamforming gain, since the thickness or beamwidth of the line and the cross-section through the line at each subcarrier determine the capacity provided to a user at every angle. To maximize the minimum capacity, we thus aim to minimize the supported \ac{AoD} range and the slope for every user. 

%% Anchor and Slope selection
To design slanted beams, we first select an \ac{AoD}-frequency \textit{anchor} for each user, or the combination of an \ac{AoD} center $\overline{\theta}_u$, \ac{AoD} range $r$, and sub-band that define each user's slanted beam. 
Since we assume user movement $\theta_u(t_i)$ is Gaussian distributed with mean and variance given in \eqref{eq:move_mean} and \eqref{eq:move_var} respectively, we use the Q-function to find $\Theta_{u,i}$, the smallest set of $\theta_u\left(t_i\right)$ with selected probability $p$.
Each user's individual center \ac{AoD} $\overline{\theta}_u$ and range $r_u$ are then the median and range of $\Theta_{u}$, the union of all $\Theta_{u,i}$. The final range $r$ used by all users is the maximum of all $r_u$.
Each \ac{UE} beam must use the same AoD-frequency slope, since \ac{TTD} arrays are, to the authors' best knowledge, incapable of creating split beams with a different slope per sub-band.
Fig. \ref{fig:aod_range} demonstrates this selection for one user, where the red dashed lines of the anchor \ac{AoD} range edges cover the maximum and minimum of the required range $r$ for $p$=97\%, shown with white lines. 
Algorithm \ref{alg:slanted_beams} describes the anchor design based on $\mathbf{g} \in \mathbb{R}^{K}$, the target directions for each subcarrier \eqref{eq:target_dir}. 
\begin{align}
    \mathbf{g}_{k_u} &= \overline{\theta}_u + \textstyle\frac{r}{2} - ru + \textstyle\frac{rU}{K}k_u, \hspace{0.3em} k_u \in \kappa_u \label{eq:target_dir} %\\
    %k_u &\in \left\{\textstyle\frac{(u-1)K}{U}+1, ..., \textstyle\frac{uK}{U}\right\}
    %\label{eq:target_idx}
\end{align}

%% Beam AWV Generation
Finally, to generate our target slanted beam, we use the \ac{JPTA} algorithm \cite{Ratnam2022_ttd_jpta} to synthesize an \ac{AWV} for a \ac{TTD} array. 
This algorithm iteratively optimizes the \acp{TTD} and phase shifts to point the beam's main lobe towards a provided target $\mathbf{g}_k, \; k \in \{1,K\}$. Effectively, the JPTA algorithm solves the following optimization problem:
\begin{align}
    \label{eq:jpta_problem}
    \max_{\mathbf{\phi}, \mathbf{\tau}}& \hspace{0.5em} \textstyle\sum_{k=1}^{K} \left|\mathbf{v}_{k}^{H}\frac{\mathbf{a}(\mathbf{g}_k,f_k)}{\|\mathbf{a}(\mathbf{g}_k,f_k)\|} \right|\\
    \mathrm{s.t.} & \hspace{0.5em} \mathbf{\phi}_n \in [-\pi, \pi], \hspace{0.5em} \mathbf{\tau}_n \in [0, \tau_{\text{max}}] \hspace{0.5em} \forall n \in\{1,...,N_t\} \nonumber
\end{align}

As \eqref{eq:jpta_problem} is non-convex, JPTA utilizes an alternating minimization with a Taylor approximation to approach a local optimum, $\phi_n^{*},\tau_n^{*}, \forall n$.
Based on our anchor design, we select a line of targeted pointing angles for each subcarrier in each user's sub-band and AoD sector. JPTA then optimizes the \ac{PS} and \ac{TTD} weights for this vector of target directions.

%% Water-filling
In this work, we assume a uniform power distribution for subcarrier water-filling. Since the true movement of the users is not known prior to beamforming or transmission, the effective frequency fading from the slanted beams is also unknown. A uniform power allocation is intuitively reasonable, as it guarantees that all \ac{AoD}s within the required range will have an equal capacity. However, a formal proof for the optimal power allocation strategy is left for future work. 

\begin{algorithm}[t] %[H]
\caption{Slanted Beams Design Algorithm}
\begin{algorithmic}
\STATE 
\STATE \textbf{Inputs:} $\widehat{\theta}_{u}\left(t_0\right), \widehat{\omega}_{u}\left(t_0\right), \widehat{\alpha}_{u}, \sigma_{\theta}, \sigma_{\omega}, \sigma_{\alpha}, \forall u \in \{1,...,U\}$
\STATE \textbf{Output:} Slanted TTD Beam $\mathbf{v}\left(f\right)$
\STATE
\STATE \textbf{select randomly} \ac{UE} sub-band assignment
\STATE $\ell = Q(1 - \frac{1-P}{2})$
\STATE \textbf{for} user $u \in [1,...,U]$
\STATE \hspace{0.5cm} \textbf{for} time step $i \in [1,...,N_s]$ 
\STATE \hspace{0.5cm}\hspace{0.5cm} Compute movement $\widehat{\theta}_{u}\left(t_i\right)$ \eqref{eq:move_mean} and $\sigma_i$ \eqref{eq:move_var}
\STATE \hspace{0.5cm}\hspace{0.5cm} $\Theta_{u,i} = \left[ \widehat{\theta}_{u}\left(t_i\right)-\sqrt{\sigma_i}\ell,  \widehat{\theta}_{u}\left(t_i\right)+\sqrt{\sigma_i}\ell \right]$
\STATE \hspace{0.5cm} $\Theta_u = \textstyle\bigcup_{i=1}^{N_s} \Theta_{u,i}$
\STATE \hspace{0.5cm} $r_u = \max\left(\Theta_u\right) - \min\left(\Theta_u\right) =$ user $u$ AoD range
\STATE \hspace{0.5cm} $\overline{\theta}_u = r_u/2 + \min\left(\Theta_u\right) =$ user $u$ AoD anchor center
\STATE $r = \max_u\left(r_u\right) =$ AoD anchor range
\STATE Select target AoDs $\mathbf{g}$ for all sub-bands using $\overline{\theta}_u, r$ in \eqref{eq:target_dir}
\STATE Solve for phases $\mathbf{\phi}$ and delays $\mathbf{\tau}$ using \textsc{JPTA} \cite{Ratnam2022_ttd_jpta} with $\mathbf{g}$ 
\STATE Create a beam $\mathbf{v}$ from $\mathbf{\phi}$ and $\mathbf{\tau}$ using \eqref{eq:ttd_bf}
\STATE \textbf{return}  $\mathbf{v}\left(f\right)$
\end{algorithmic}
\label{alg:slanted_beams}
\end{algorithm}

% \subsection{User Sub-band Assignment}
%TODO LATER IF INTERESTING RESULTS COME UP

\subsection{Baseline Analog Beamforming}
\label{subsec:beams_baselines}

Several analog beamforming approaches can be used to attempt to provide reliable coverage to multiple users through distinct \ac{OFDMA} sub-bands. Table \ref{table:beams} summarizes all the included reference beams. 
In this work, we provide two primary baseline beam designs: stepped \ac{TTD} beams and rainbow beams. 
Stepped \ac{TTD} beams \cite{Ratnam2022_ttd_jpta, Zhao2024_ttd_step} split non-dispersive beams towards each user's initial \ac{AoD} estimate $\widehat{\theta}_u\left(t_{0}\right)$ over their entire sub-band. Since stepped beams are equivalent to slanted beams with $r=0$, we use JPTA to design the \acp{AWV}. 
Dispersive rainbow beams \cite{Yan2019_ttd_rainbow} spread beam pointing directions across the entire \ac{AoD} range over the entire bandwidth, presenting the asymptotic equivalent of slanted beams as $U \to \infty$ and user sub-band associations are sorted by \ac{AoD}.

We also include three baseline designs that are not practical for this system but demonstrate the benefits of slanted beams.  
The two \textit{genie} baselines require unrealistic \textit{genie} knowledge of $\theta_u\left(t_i\right)$, the true \ac{AoD}s of all \ac{UE}s at all time steps. The stepped genie points a \ac{TTD} stepped beam to all $\theta_u(t_i)$ instead of the initial estimates $\widehat{\theta}_u(t_0)$. This beam provides among the best performance available from a \ac{TTD} array, though we will later show stepped beams may be sub-optimal for large numbers of users. The optimal genie uses a digital array beamformer to point perfect sub-band beams to all  $\theta_u(t_i)$, providing maximum performance for any array of this size. Though a digital array could beamform the entire bandwidth to every user independently, we assume that the users must only use their own sub-band for a fair comparison with the analog beams.
Finally, \ac{QPD} \cite{Sayidmarie2013_wide_QPD} beams provide a heuristic method to efficiently create single-lobe wide beams \acp{PA}. As a single-user beam design, QPD beams represent a lower limit for robust beamformer design.

\begin{table}
\caption{Comparison of Beamforming Methods}
\label{table:beams}
\setlength{\tabcolsep}{3pt}
\makegapedcells
\begin{tabular}{|p{55pt}|p{25pt}|p{25pt}|p{112pt}|}  %{|p{145pt}|p{85pt}|} %
%{|p{25pt}|p{75pt}|p{115pt}|} %
\hline
\textbf{Beam}& \textbf{Array}& \textbf{Width}& \textbf{AWV or Phases and Delays} \\
\hline
\hline
\textit{Slanted} & \textit{TTD} & \textit{Wide} & See Algorithm \ref{alg:slanted_beams} \\
\hline
Rainbow & TTD & Wide & $\mathbf{\tau}_n = -n/(2B)$, \\[1ex]
&&& $\mathbf{\phi}_n = 2\pi n f_c/W$ \\[1ex]
\hline
Stepped & TTD & Narrow & JPTA with $\mathbf{g}_{k_u} = \widehat{\theta}_u\left(t_{0}\right)$ \\[1ex]
\hline
Stepped Genie & TTD & Narrow & JPTA with $\mathbf{g}_{k_u}(t_i) = \theta_u(t_i)$ \\[1ex]
\hline
Optimal Genie & Digital & Narrow & $\mathbf{v}_n\left(t_i, f_k\right) = N_t^{-1/2} \times$ \\[1ex]
&&& $\exp{ j2\pi d \frac{f_k}{c} \sin\theta_{u}\left(t_i\right) } $ \\[1ex]
QPD ($\Phi_{m} = \pi$) & PA & Wide & $\mathbf{\phi}_n = \textstyle\frac{2\pi d f_c}{c} \sin(\widehat{\theta}_1\left(t_{0}\right)) + \phi_{q, n}$ \\
&&& $\phi_{q, n} = 4\Phi_m \left(\frac{2 n - N_t - 1}{2(N_t + 1)} \right)^2$ \\[1ex]
\hline
\end{tabular}
\vspace{-3mm}
\end{table}

\section{Simulation Results}
\label{sec:results}

Our simulation system parameters, unless otherwise specified for parameter sweeps, are shown in Table \ref{table:sims} for our 100 \ac{MC} trials comparing beamforming performance. These parameters were selected to provide realistic mobility and error. The initial \ac{AoD} error variance is based on sub-beamwidth \ac{BA} accuracy for the 32-element array's $~3.2\degree$ beams. 
For a \ac{UE} 3m from the boresight of the \ac{BS} array, the 3.2 m/s maximum velocities found in experimental data \cite{Marinsek2024_urllc_vr} corresponds to 47 deg/s angular velocity.
We select the maximum 5G-NR tracking frame length, $T=160$ ms, saving 7/8 \ac{BA}s compared to a typical 20 ms 5G-NR SSB period.
We target high-resolution \ac{XR} requirements \cite{Alriksson2023_xr_reqs} including 100 Mbps \ac{DL} per-user data rate and 99.99\% end-to-end reliability.

\begin{table}
\caption{Simulation Parameters}
\label{table:sims}
\setlength{\tabcolsep}{3pt}
\begin{tabular}{|p{165pt}|p{65pt}|}  %{|p{145pt}|p{85pt}|} %
%{|p{25pt}|p{75pt}|p{115pt}|} %
\hline
% \multicolumn{2}{|c|}{\textbf{Simulation System Parameters}}\\
% \hline
\textbf{Parameter}& \textbf{Value} \\
\hline
Center frequency ($f_c$) & 60 GHz \\
Bandwidth ($W$) & 2 GHz \\
Number of subcarriers ($K$) / Bandwidth ($w_{sc}$) & 1200 / 1.67 MHz \\
Number of Tx antennas ($N_t$) & 32 \\
Antenna spacing ($d$) & $0.5\lambda$ \\
\ac{SNR} w/o beamforming & -10 dB \\
Initial AoD estimate range (for $\widehat{\theta}_{u}\left(t_{0}\right)$) & $[-45\degree, 45\degree]$ \\
Minimum initial \ac{UE} AoD spacing & 10\degree \\
Number of \ac{UE}s ($U$) & 3 \\
% Number of Monte Carlo trials & 100 \\
\hline
\hline
\ac{BA} frame time ($T$) & 160 ms \\
Step time ($\Delta t$; for $R = 100$ steps) & 1.6 ms \\
Coverage probability parameter ($p$) & 97\% \\
Initial AoD error variance ($\sigma_{\theta}$) & 2 deg\textsuperscript{2} \\
Mean initial angular velocity range ($\widehat{\omega}_{u}\left(t_{0}\right)$) & [0, 80] deg/s \\ 
Initial angular velocity variance ($\sigma_{\omega}$) & 10 deg\textsuperscript{2}/s\textsuperscript{2} \\
Mean constant angular acceleration ($\widehat{\alpha}$) & 0 deg/s\textsuperscript{2} \\
Constant angular acceleration variance ($\sigma_{\alpha}$) & 5 deg\textsuperscript{2}/s\textsuperscript{4} \\
\hline
\end{tabular}
\end{table}

\begin{figure*}[t]
    \vspace{0mm}
    \centering
    % left bottom right top
    \includegraphics[width = 0.94\linewidth,trim=65 10 25 10,clip]{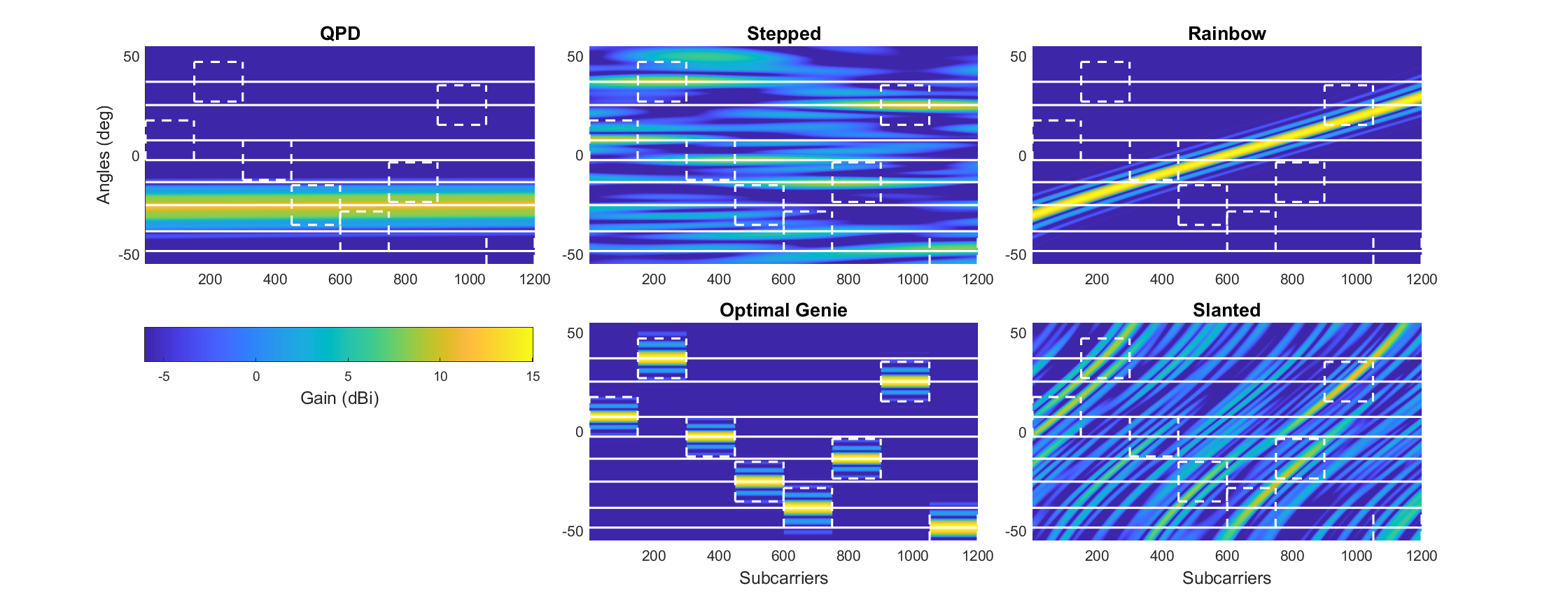}
    \vspace{-3mm}
    \caption{Example comparison of all beam designs with $U=8$, with the user anchors marked in white.}
    \label{fig:test2_beamComp_8users}
    \vspace{-7mm}
\end{figure*}

Before discussing the capacity performance results, we provide example antenna patterns from the gamut of beam designs. Fig. \ref{fig:test2_beamComp_8users} shows the gain from the 6 different beams as heatmaps over the \ac{AoD}-frequency domain. The beams were designed for a scenario with a random \ac{AoD} for each of 8 users. For this plot, the genie beams only show the pattern designed for the center AoD, as these beams adjust for every true \ac{AoD}. Compared to the 3 user example in Fig. \ref{fig:system}, we see many additional side lobes as diagonal striations in the heatmap. This is expected, as the 32-element array averages only 4 antennas per user to service the 8 users simultaneously.

\subsection{Minimum Capacity with Random AoD Offsets}
\label{subsec:results_offsets}

\begin{figure}[t]
    \vspace{0mm}
    \centering
    % left bottom right top
    \includegraphics[width = 0.99\linewidth,trim=10 0 10 0,clip]{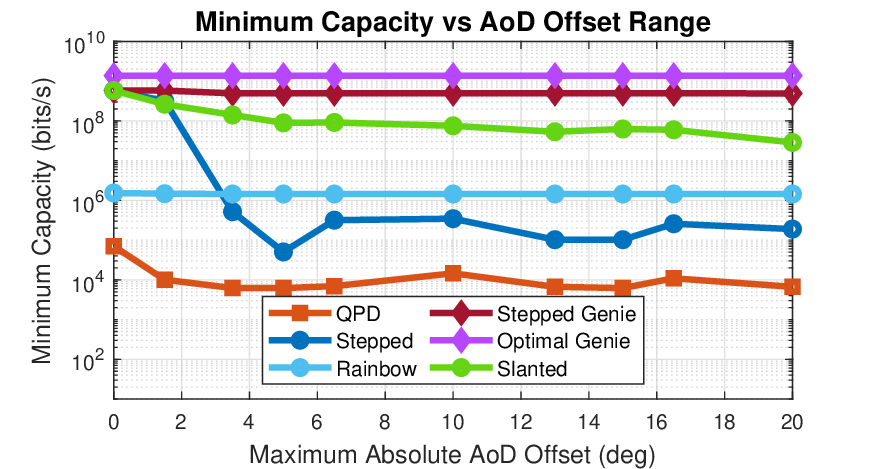}
    \vspace{-7mm}
    \caption{Robustness of capacity to larger maximum AoD offsets.}
    \label{fig:test3_aod_range}
    \vspace{-5mm}
\end{figure}

Our first set of simulation results demonstrate slanted beams' robustness to random \ac{UE} \ac{AoD} errors from \ac{BA} error or unmeasured movement and uncover the circumstances where slanted beams are most beneficial. Except in Fig. \ref{fig:test3_aod_range}, the beams in this section were designed with a range $r=20\degree$. We computed the capacity per user over a grid of 100 uniformly-spaced angle offsets centered on the random \ac{UE} \ac{AoD}s for each \ac{MC} trial. We use the \textit{minimum capacity} metric, $M=\min_{u, i} C_u\left(t_i, \mathbf{v}\right)$, to measure the minimum total capacity summed over sub-band subcarriers among all users and tested angle offsets. 
Fig. \ref{fig:test3_aod_range} shows $M$ for each beam design as the range of \ac{AoD} offsets, $\theta_u - \widehat{\theta}_{u}$, increases from $\pm0\degree$ to $\pm20\degree$. Slanted beams operate as designed; they show little degradation in $M$ with larger offsets and provide the best result besides genie methods. As we will see in all results, the stepped genie beams perform well but are always worse than the digital genie beams. Rainbow beams provide consistent $M$ at all offset ranges as they are independent of user \ac{AoD}s. Unsurprisingly, $M$ for stepped and \ac{QPD} beams is the lowest for large angle offsets, as stepped beams narrowly focus on $\widehat{\theta}_{u}$ while \ac{QPD} beams are only designed for one of three users.

\begin{figure}[t]
    \vspace{0mm}
    \centering
    % left bottom right top
    \includegraphics[width = 0.99\linewidth,trim=10 0 10 0,clip]{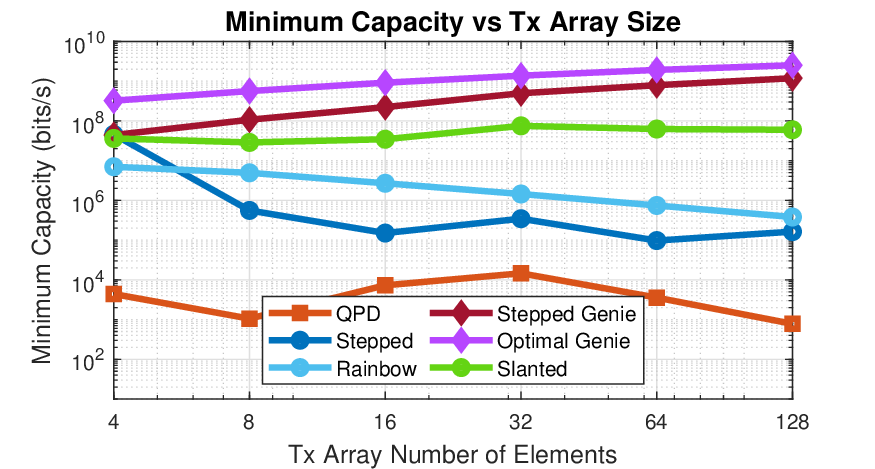}
    \vspace{-7mm}
    \caption{Impact of Tx array size on robustness to AoD offsets.}
    \label{fig:test3_txSize}
    \vspace{-5mm}
\end{figure}

We compare the beam designs' minimum capacity for random AoD offsets and varied \ac{BS} array sizes in Fig. \ref{fig:test3_txSize}. Slanted beams provide consistent $M$ over array sizes, since larger $N_t$ provides higher peak gain but narrower supported bandwidth. The frequency cross-section of the angled linear beam pattern governs the effective bandwidth for a given \ac{AoD}. 
Rainbow beams degrade $M$ by orders-of-magnitude for larger $N_t$, as the narrower beams reduce their effective angle-frequency coverage. 
Genie beams have the true \acp{AoD}, do not follow this tradeoff, and improve with larger $N_t$. 
All other beams show even lower $M$, making slanted beams the only practical design to handle \ac{AoD} uncertainty with large $N_t$. 

\begin{figure}[t]
    \vspace{0mm}
    \centering
    % left bottom right top
    \includegraphics[width = 0.99\linewidth,trim=10 0 10 0,clip]{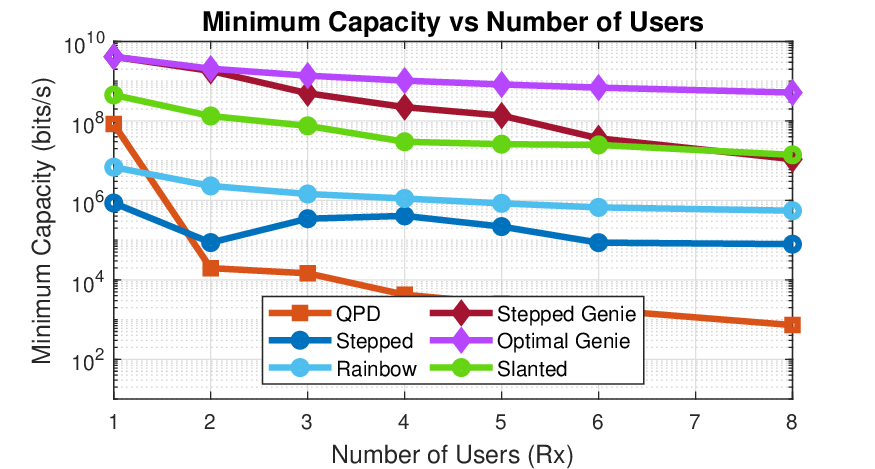}
    \vspace{-7mm}
    \caption{Robust beamforming performance vs number of users.}
    \label{fig:test3_n_users}
    \vspace{-5mm}
\end{figure}

Fig. \ref{fig:test3_n_users} demonstrates how well the beam designs scale with more simultaneous users. Slanted beams are consistent again, even outperforming the genie TTD stepped beams when $U=8$. As illustrated in Fig. \ref{fig:test2_beamComp_8users}, analog arrays struggle to produce beams with multiple distinct main lobes for many users. With our simulation parameters, $U\geq8$ reduces stepped beams' gain to where they are no longer optimal for \ac{OFDMA}. 
We also found that the genie information is less relevant for TTD stepped beams with larger $U$. Stepped beams smear more across AoDs reducing the stepped genie performance. While the optimal digital beamformer also degrades with larger $U$, as the bandwidth per user decreases, the stepped genie beams degrade much faster.

\subsection{Minimum Capacity with Random Movement}
\label{subsec:results_movement}

\begin{figure}[t]
    \vspace{0mm}
    \centering
    % left bottom right top
    \includegraphics[width = 0.99\linewidth,trim=10 0 10 0,clip]{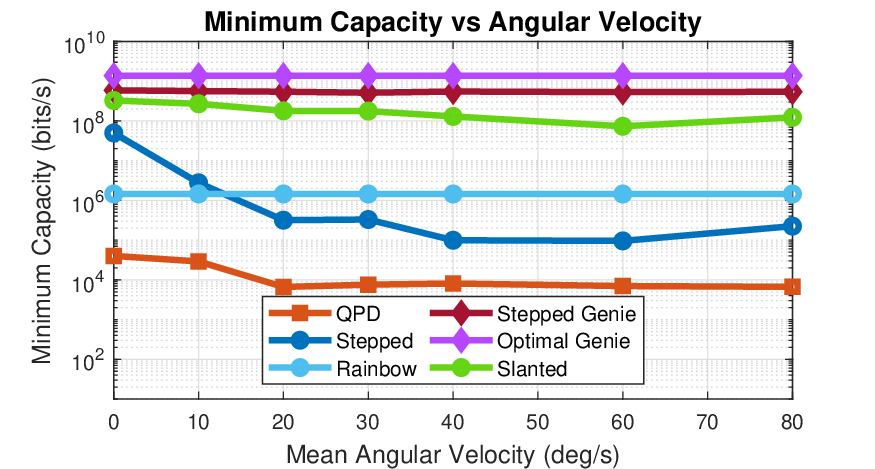}
    \vspace{-7mm}
    \caption{Minimum capacity with increased UE velocity.}
    \label{fig:test4_velocity}
    \vspace{-5mm}
\end{figure}

\begin{figure}[t]
    \vspace{0mm}
    \centering
    % left bottom right top
    \includegraphics[width = 0.99\linewidth,trim=0 0 25 0,clip]{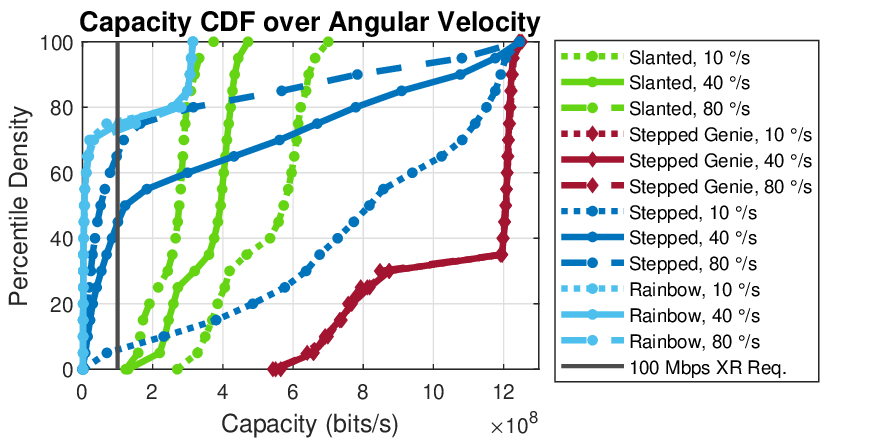}
    \vspace{-7mm}
    \caption{Capacity CDFs for three UE velocities.}
    \label{fig:test4_cdf}
    \vspace{-5mm}
\end{figure}

Finally, we show the performance of slanted beams with \ac{UE} movement. In Fig. \ref{fig:test4_velocity}, we sweep mean initial angular velocities $\widehat{\omega}_u\left(t_{0}\right)$ to test faster movement. The minimum capacity results are similar to Fig. \ref{fig:test3_aod_range}, since the random motion introduces a different \ac{AoD} offset at each time step, larger velocities increase the range $r$, and thus beam robustness to larger $\widehat{\omega}_u\left(t_{0}\right)$ is  robustness to larger \ac{AoD} offsets. As in Fig. \ref{fig:test3_aod_range}, slanted beams provided one to two orders-of-magnitude better $M$ than the next best beams, rainbow beams. %, when $\omega_u > 10$. 
Slanted beams also provide higher capacity in more scenarios than the minimum. The capacity \acp{CDF} in Fig. \ref{fig:test4_cdf} summarize all \ac{MC} trials, with its x-intercepts presenting select results from \ref{fig:test4_velocity}. For the three mean velocities shown, the proposed slanted beams \textit{always} met the target 100 Mbps capacity, while the stepped and rainbow beams did not in 5\%-75\% of cases. At high velocities, slanted beams also provided higher capacity than stepped beams for most trials; 65\% and 80\% of cases for 40 deg/s and 80 deg/s respectively.

\section{Conclusion and Future Work}
\label{sec:conclusion}

In this paper, we propose \ac{TTD} array based slanted beams, a novel method to create wideband, downlink beamforming that is robust to randomized user movement. The proposed beam design algorithm provides a simple tradeoff between beam frequency selectivity and wider angular coverage. 
Our simulation results show that slanted beams provide more reliable coverage for multiple users over a specified angular range or for random movement. Slanted beams provided up to two orders-of-magnitude better minimum capacity compared to baseline beam designs.
Future work based on this beamforming method includes multi-frame tracking and protocol augmentations to improve continuous tracking.

\bibliographystyle{IEEEtran}
\bibliography{IEEEabrv,references}

\end{document}